\documentclass{moriond}

\usepackage{amsmath}

\def\be{\begin{equation}}
\def\ee{\end{equation}}
\def\bea{\begin{eqnarray}}
\def\eea{\end{eqnarray}}

\newcommand{\as}{\alpha_s}
\newcommand{\Photo}{}

\begin{document}
\title{Precision physics in jet processes}

\author{Fr\'ed\'eric A. Dreyer}

\address{Sorbonne Universit\'es, UPMC Univ Paris 06, UMR 7589,
  LPTHE, F-75005, Paris, France\\
  CNRS, UMR 7589, LPTHE, F-75005, Paris, France\\
  CERN, PH-TH, CH-1211 Geneva 23, Switzerland}

\maketitle\abstracts{ 
  We examine properties of small-radius jets,
  focusing on phenomenological applications to the inclusive jet
  spectrum.
  We match fixed-order calculations with the leading-logarithmic
  resummation of the jet radius (LL$_R$), and propose a new prescription
  to evaluate theoretical uncertainties for next-to-leading order
  (NLO) predictions.
  We also examine the $R$-dependent next-to-next-to-leading order (NNLO)
  corrections, and include them in our calculation.
  We discuss hadronisation corrections, which are derived from Monte
  Carlo generators.
  Finally, we assemble these elements and compare the ratio of the
  inclusive jet spectra at two $R$ values obtained from our matched
  (N)NLO+LL$_R$ predictions to data from ALICE and ATLAS, finding
  improved agreement.}

\section{Introduction}
Achieving highly precise predictions is an essential goal of modern
collider physics.
It is important in numerous contexts, notably: in Higgs physics, where
accurate determination of couplings are now one of the main goals; in
PDF extractions, whose uncertainties feed back into all other
theoretical predictions; and in the determination of electroweak
parameters.
A large number of analyses at the LHC rely on the use of jets, and as
such, an understanding of how limits on precision could be pushed in
such processes would bring valuable insights.

We consider an archetypal hadron-collider jet observable, the
inclusive jet spectrum, which can provide a useful case study with
both experimental and theoretical challenges.
We aim to investigate the $R$-dependence of jet spectra, focusing
particularly on the small-radius limit.

\section{Small-radius resummation}
Building on previous work treating the resummation of leading
logarithmic small-radius terms~\cite{Dasgupta:2014yra}, it is
straightforward to express the small-$R$ resummed inclusive
``microjet'' spectrum from the convolution of the inclusive microjet
fragmentation function,
$f^\text{incl}_{\text{jet}/k}(p_t/p'_t,t)$, with the leading order
(LO) inclusive spectrum of partons with transverse momentum $p_t'$ and
flavour $k$, $\frac{d\sigma^{(k)}}{dp'_t}$
\begin{equation}
  \label{eq:LLR-master}
  \sigma^{\text{LL}_R}(p_t,R) \equiv
  \frac{d\sigma_\text{jet}^{\text{LL}_R}}{dp_t} =
  \sum_k\int_{p_t}\frac{dp'_t}{p'_t}\,
  f^\text{incl}_{\text{jet}/k}\left( \frac{p_t}{p'_t},t(R,R_0,\mu_R) \right) \,
  \frac{d\sigma^{(k)}}{dp'_t}\,.
\end{equation}
Here we use an evolution variable $t$, defined by
\begin{equation}
  \label{eq:t}
  t(R, R_0, \mu_R) = \int_{R^2}^{R_0^2} \frac{d\theta^2}{\theta^2}  
  \frac{\as(\mu_R\,
    \theta/R_0)}{2\pi}\,,\quad
  b_0 = \frac{11 C_A - 4 T_R n_f}{6}\,,
\end{equation}
where $R_0\sim 1$ is an angular scale at which the small-$R$
approximation becomes valid.

\begin{figure}[ht]
  \centering
  \includegraphics[width=0.42\textwidth]{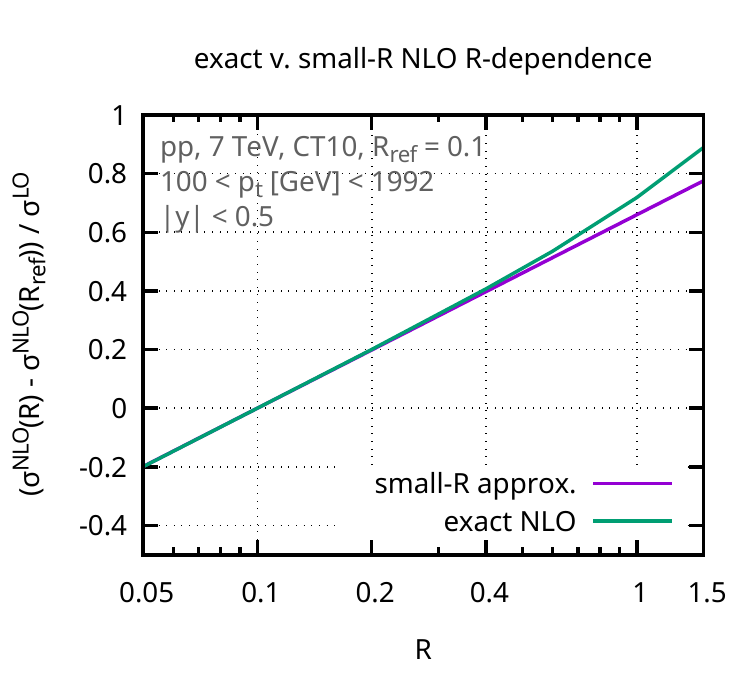}~
  \includegraphics[width=0.42\textwidth]{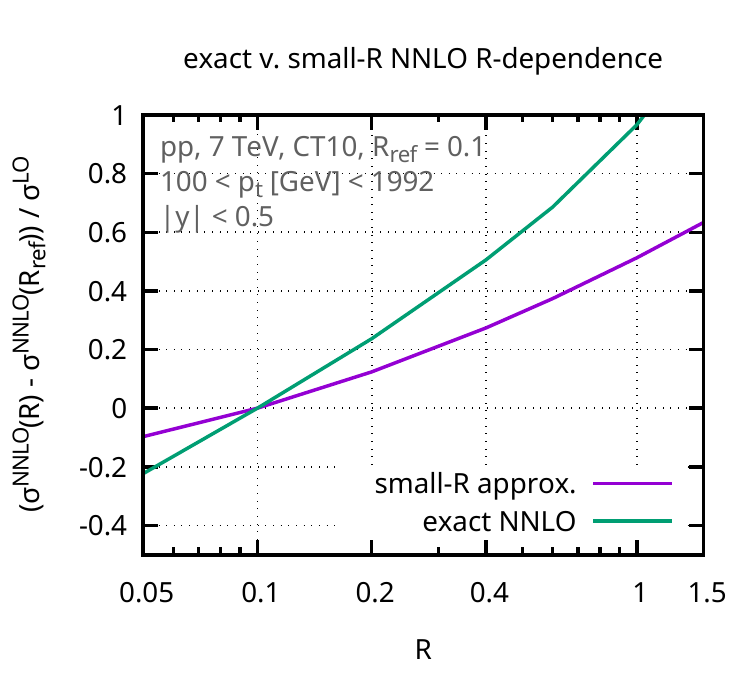}
  \caption{The exact and small-$R$ approximation for the $R$ dependence
    of the cross section, at NLO (left) and NNLO (right).}
  \label{fig:small-R-validity}
\end{figure}

To verify the accuracy of the small-$R$ approximation, we can compare
the difference between $R$ values obtained from a fixed-order
calculation and the corresponding result obtained from the expansion
of the resummed result given by Eq.~(\ref{eq:LLR-master}).
Fig.~\ref{fig:small-R-validity} (left) shows
$\tfrac1{\sigma^\text{LO}}\big(\sigma^\text{NLO}(R)-\sigma^\text{NLO}(R_\text{ref})\big)$
as a function of $R$ for both the small-$R$ expansion and the exact
fixed-order calculation (obtained with
\texttt{NLOJet++}~\cite{Nagy:2003tz}), taking $R_\text{ref}=0.1$.
We observe that the small-$R$ approximation works very well for values
of the jet radius observing $R\leq 0.6$.

Furthermore, to examine subleading terms beyond NLO, we take a NLO 3-jet
calculation, using the fact that
\begin{equation}
  \label{eq:nlo3j-nnlo}
  \sigma^{\text{NNLO}}(R)-\sigma^{\text{NNLO}}(R_\text{ref})
  = \sigma^{\text{NLO}_{3j}}(R) - \sigma^{\text{NLO}_{3j}}(R_\text{ref})\,.
\end{equation}
Comparing again the exact result with the expansion of the resummed
spectrum, we can see in Fig.~\ref{fig:small-R-validity} (right) that
there are important subleading contributions of the form
$\as^n \ln^{n-1}R$.
We will include a subset of these terms by matching the LL$_R$
resummation with the exact $R$ dependence up to NNLO.

\section{Matching to fixed order}
To obtain phenomenological predictions, we combine the LL$_R$
resummation with fixed-order results.
We achieve this by adopting a multiplicative matching scheme, given by
\begin{equation}
  \label{eq:multiplicative-matching}
  \sigma^{\text{NLO}+\text{LL}_R} = 
  \left(\sigma_0 + \sigma_1(R_0) \right)
  \times 
  \left[\frac{\sigma^{\text{LL}_R}(R)}{\sigma_0}
  \times \left(1 + \frac{\sigma_1(R)- \sigma_1(R_0) - \sigma^{\text{LL}_R}_1(R)}{
      \sigma_0}\right)\right],
\end{equation}
where $\sigma_i$ is the contribution of order $\as^{2+i}$ to the
inclusive jet cross section, and the superscript LL$_R$ signals
predictions obtained from the small-$R$ approximation.
Because the two terms in separate brackets in
Eq.~(\ref{eq:multiplicative-matching}) lead to K-factors going in
opposite directions, there is a partial cancellation of higher order effects.
This leads to unphysical cancellations of the scale uncertainties for
certain values of the jet radius.
Therefore, we propose an alternative method to estimate missing higher
orders uncertainties, which we obtain by evaluating the scale
dependence independently in each term, and adding the resulting
uncertainties in quadrature.

Furthermore, because of the importance of subleading $\as^2\ln R$
terms highlighted in Fig.~\ref{fig:small-R-validity} (right), it is
important to include them.
To this end, we construct a stand-in for the full NNLO result, denoted
NNLO$_R$, which contains the complete $R$ dependence
\begin{equation}
  \label{eq:sigma-nnlo}
  \sigma^{\text{NNLO}_R}(R,R_m) \equiv \sigma_0 + \sigma_1(R) + [\sigma_2(R) -
  \sigma_2(R_m)].
\end{equation}
Here $R_m$ is an arbitrary angular scale, which we will take as
$R_m=1$.
We then extend the NLO matching scheme described in
Eq.~(\ref{eq:multiplicative-matching}) up to NNLO to obtain
matched NNLO$_R$+LL$_R$ predictions~\cite{Dasgupta:2016bnd}.

\section{Non-perturbative effects}
In order to compare our predictions with data, it is important to
consider the impact of non-perturbative effects.
The two main sources of non-perturbative corrections are:
hadronisation, which is the transition from parton-level to
hadron-level; and underlying event (UE), which corresponds to multiple
interactions of the partons in the colliding protons.

\begin{figure}[ht]
  \centering
  \includegraphics[width=0.45\textwidth]{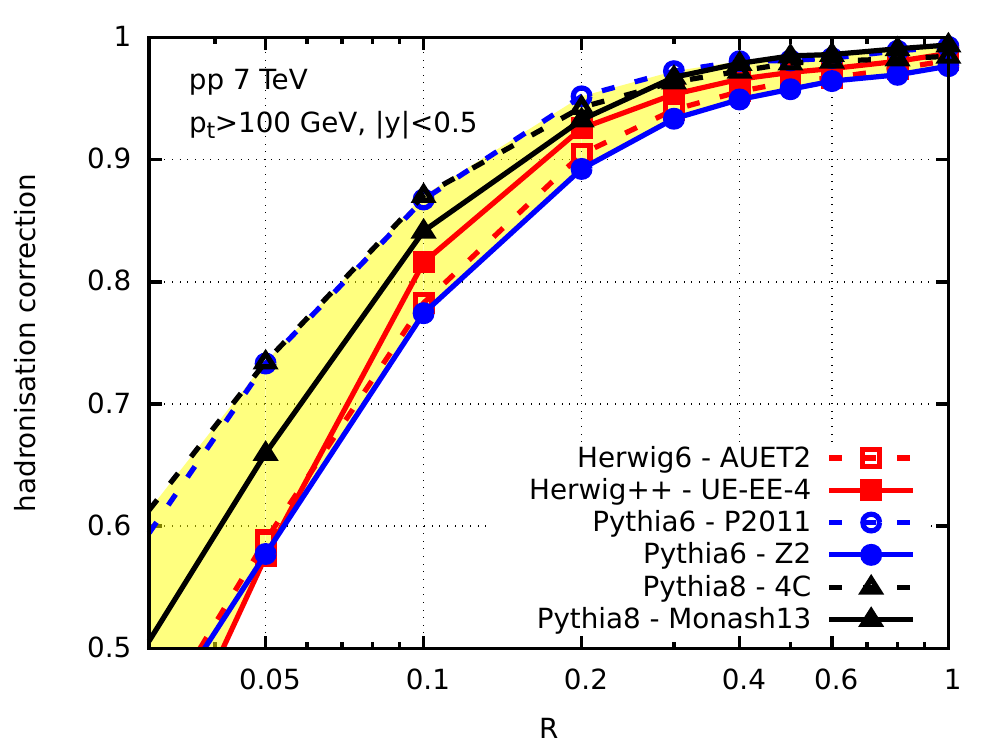}~
  \includegraphics[width=0.45\textwidth]{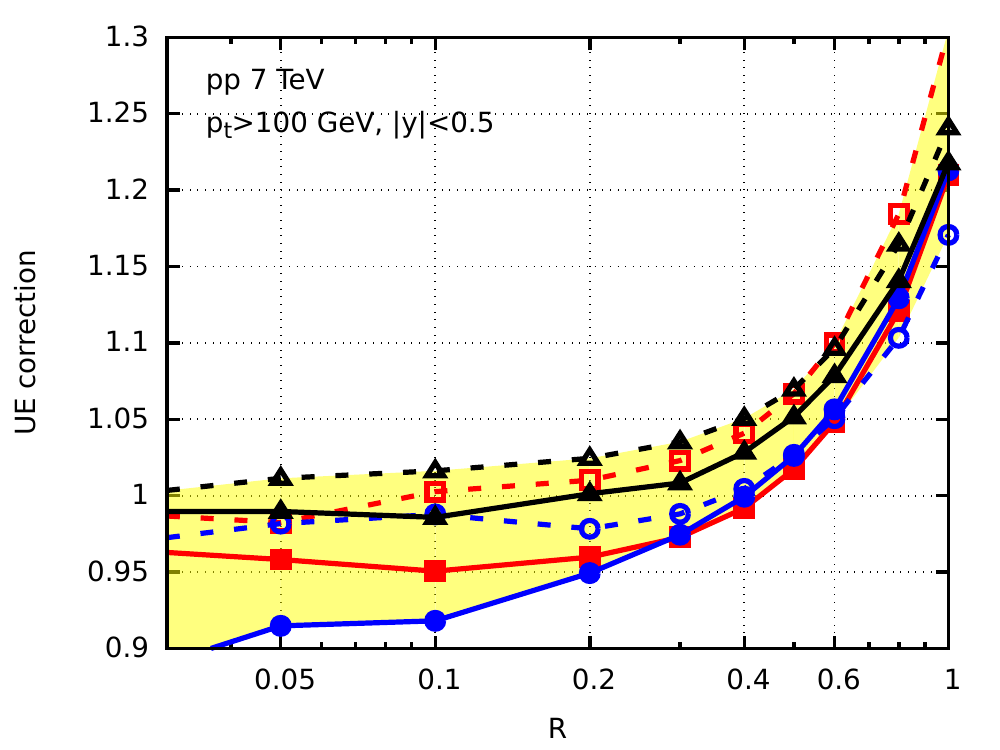}
  \caption{Non-perturbative corrections factors derived from Monte
    Carlo event generators, for hadronisation (left) and underlying
    event (right).}
  \label{fig:np-corrections}
\end{figure}

Their dependence on the jet radius is very different, therefore it is
useful to examine them separately: hadronisation is enhanced as $1/R$
at small radii, while the shift in jet $p_t$ from UE scales as $R^2$.
The correction factors derived from 6 different Monte Carlo tunes is
shown in Fig.~\ref{fig:np-corrections} as a function of $R$, both
for hadronisation (left) and UE (right).

We include non-perturbative effects by rescaling the perturbative
spectra with factors derived from the average of several Monte Carlo
tunes.

\section{Comparison to data}
We can now compare our predictions with existing inclusive jet data
from the ALICE and ATLAS experiments.

\begin{figure}[ht]
  \centering
  \includegraphics[width=0.45\textwidth]{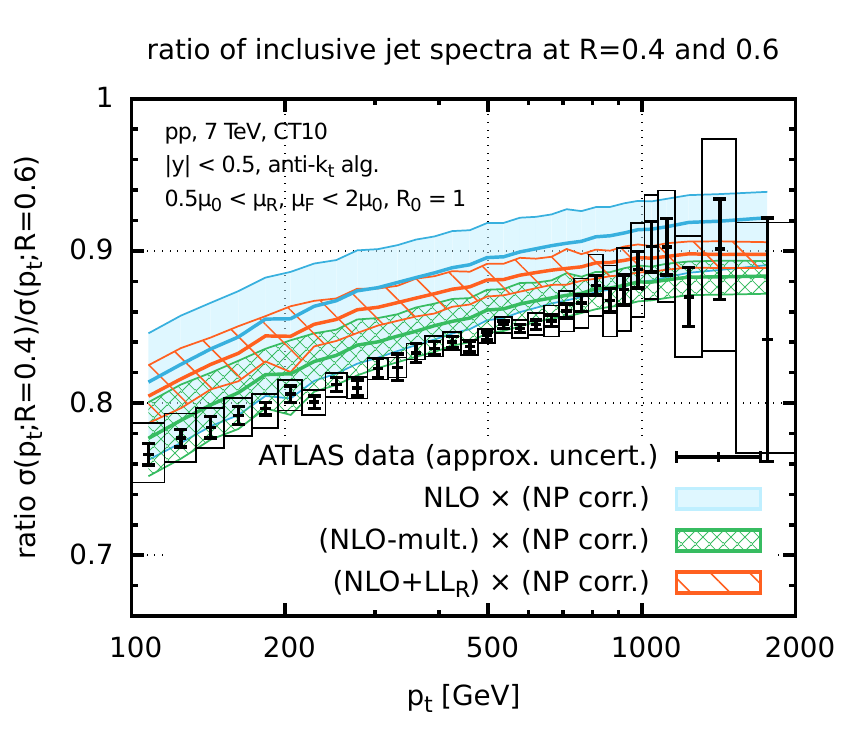}~
  \includegraphics[width=0.45\textwidth]{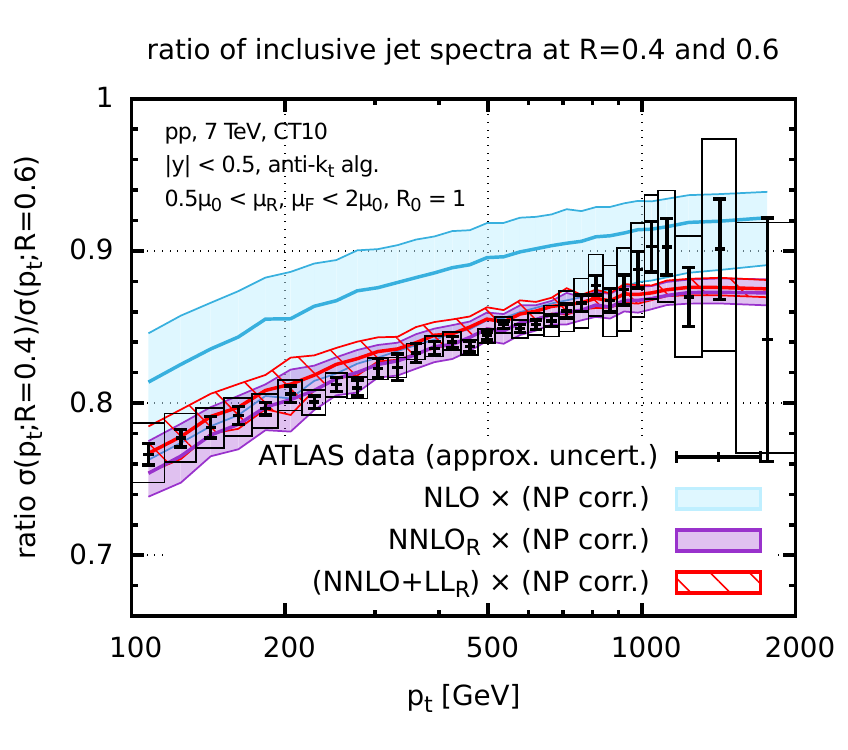}
  \caption{Comparison of theoretical predictions with data from the
    ATLAS collaboration~\cite{Aad:2014vwa}.}
  \label{fig:atlas-data}
\end{figure}

The ATLAS data~\cite{Aad:2014vwa} is at centre-of-mass energy
$\sqrt{s} = 7$ TeV, with two values of the jet radius: $R=0.4$ and
$R=0.6$.
To best evaluate the compatibility of our predictions with the
experimental data, we take the ratio of the inclusive jet spectra at
the two $R$ values.
This allows us to study directly the $R$ dependence, as a number of
experimental and theoretical uncertainties cancel in the ratio.
In Fig.~\ref{fig:atlas-data}, we show the result of this comparison.
Here we observe a much better agreement of the experimental data with
the NNLO$_{(R)}$ based predictions compared with the plain NLO.

\begin{figure}[ht]
  \centering
  \includegraphics[width=0.44\textwidth]{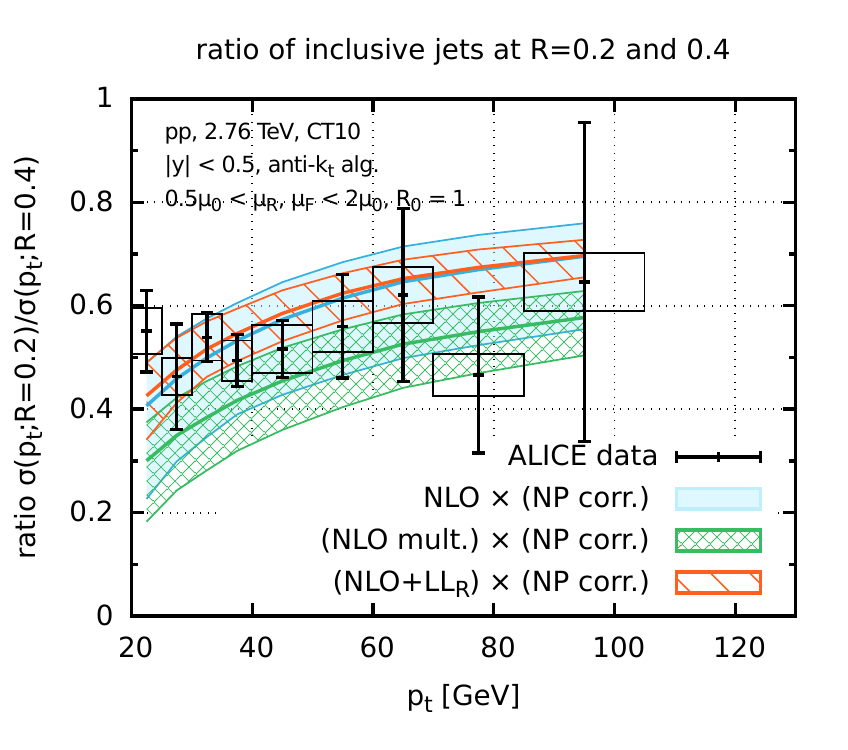}~
  \includegraphics[width=0.44\textwidth]{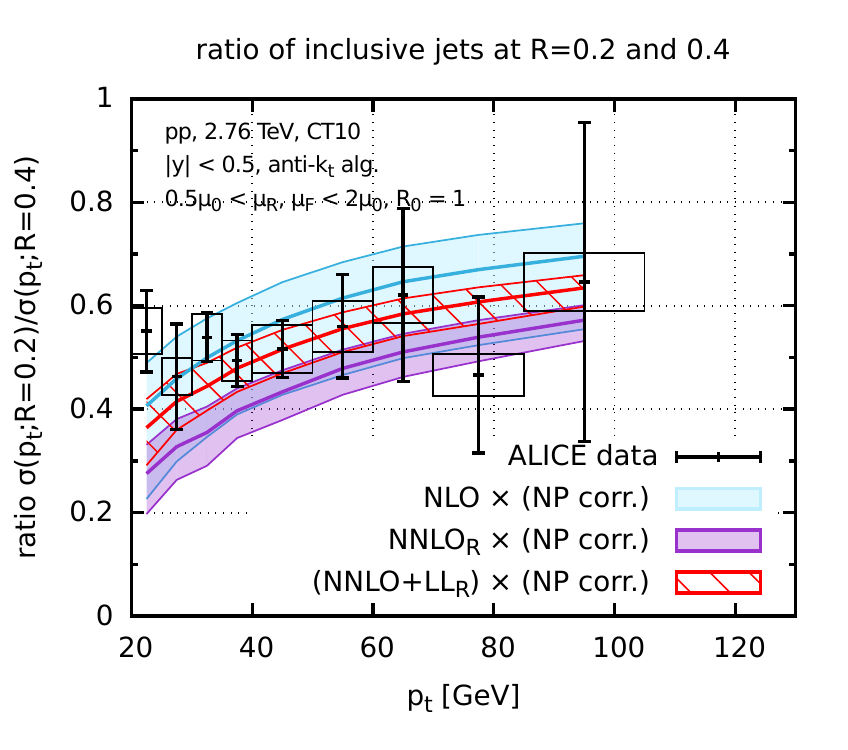}
  \caption{Comparison of theoretical predictions with data from the
    ALICE collaboration~\cite{Abelev:2013fn}.}
  \label{fig:alice-data}
\end{figure}

We also compare our results with inclusive jet data from the ALICE
collaboration~\cite{Abelev:2013fn}, taken at $\sqrt{s}=2.76$ TeV with
$R=0.2$ and $R=0.4$.
Taking again the ratio of these spectra, we show a comparison with our
theoretical predictions in Fig.~\ref{fig:alice-data}.
We can see substantial effects beyond NLO from the resummation and the
NNLO terms, with the NNLO+LL$_R$ seemingly providing the best match
for the experimental data.

\section{Conclusion}
We provided a detailed study of small-$R$ effects in inclusive jet
spectra, considering $R$-dependent contributions up to NNLO which
where matched with the LL$_R$ resummation.
We studied non-perturbative effects and included them as corrections
factors for the comparison with experimental data.
Our work suggests that using multiple $R$ values can provide a
powerful handle on systematic uncertainties.
As such, we encourage experimental collaboration to use at least three
different $R$ values for their jet measurements: $R=0.2-0.3$, where UE
is suppressed, $R=0.4$, and $R=0.6-0.7$ where hadronisation is
suppressed.
The computer code used for this study, as well as additional figures,
are available online~\cite{OnlineTool}.

\section*{Acknowledgements}
This work has been done in collaboration with Mrinal Dasgupta, Gavin
Salam and Gregory Soyez, and was supported by the ILP LABEX
(ANR-10-LABX-63) supported by French state funds managed by the ANR
within the Investissements d'Avenir programme under reference
ANR-11-IDEX-0004-02.

\end{document}